\documentclass[conference]{IEEEtran}
\IEEEoverridecommandlockouts
\usepackage{cite}
\usepackage{amsmath,amssymb,amsfonts}
\usepackage{algorithmic}
\usepackage{graphicx}
\usepackage{textcomp}
\usepackage{xcolor}
\usepackage{enumitem}
\usepackage{hyperref}
\usepackage{cleveref}
\usepackage{tikz}

\def\BibTeX{{\rm B\kern-.05em{\sc i\kern-.025em b}\kern-.08em
    T\kern-.1667em\lower.7ex\hbox{E}\kern-.125emX}}

\newcommand \copyrighttext {
  \footnotesize \textcopyright 2022 IEEE. Personal use of this material is permitted. Permission from IEEE must be obtained for all other uses, in any current or future media, including reprinting/republishing this material for advertising or promotional purposes, creating new collective works, for resale or redistribution to servers or lists, or reuse of any copyrighted component of this work in other works. DOI: \href{https://doi.org/10.1109/RE54965.2022.00041}{10.1109/RE54965.2022.00041}.
}

\newcommand\copyrightnotice{
    \begin{tikzpicture}[remember picture,overlay]
    \node[anchor=south,yshift=10pt] at (current page.south) {\fbox{\parbox{\dimexpr\textwidth-\fboxsep-\fboxrule\relax}{\copyrighttext}}};
    \end{tikzpicture}
}
    
\begin{document}

\title{A Live Extensible Ontology of Quality Factors for Textual Requirements}

\author{
    \IEEEauthorblockN{
        1\textsuperscript{st} Julian Frattini \\
        4\textsuperscript{th} Michael Unterkalmsteiner \\ 
        5\textsuperscript{th} Daniel Mendez*, 
        6\textsuperscript{th} Davide Fucci}
    \IEEEauthorblockA{\textit{Blekinge Institute of Technology}\\
        Karlskrona, Sweden \\
        \{firstname\}.\{lastname\}@bth.se}
    \and
    \IEEEauthorblockN{
        2\textsuperscript{nd} Lloyd Montgomery}
    \IEEEauthorblockA{\textit{University of Hamburg}\\
        Hamburg, Germany \\
        lloyd.montgomery@uni-hamburg.de}
    \and
    \IEEEauthorblockN{
        3\textsuperscript{rd} Jannik Fischbach}
    \IEEEauthorblockA{\textit{Netlight Consulting GmbH and *fortiss GmbH}\\
        Munich, Germany \\
        jannik.fischbach@netlight.com}
}

\maketitle

\copyrightnotice

\begin{abstract}
Quality factors like passive voice or sentence length are commonly used in research and practice to evaluate the quality of natural language requirements since they indicate defects in requirements artifacts that potentially propagate to later stages in the development life cycle. 
However, as a research community, we still lack a holistic perspective on quality factors.
This inhibits not only a comprehensive understanding of the existing body of knowledge but also the effective use and evolution of these factors. 
To this end, we propose an ontology of quality factors for textual requirements, which includes (1) a structure framing quality factors and related elements and (2) a central repository and web interface making these factors publicly accessible and usable. 
We contribute the first version of both by applying a rigorous ontology development method to 105 eligible primary studies and construct a first version of the repository and interface.
We illustrate the usability of the ontology and invite fellow researchers to a joint community effort to complete and maintain this knowledge repository. We envision our ontology to reflect the community's harmonized perception of requirements quality factors, guide reporting of new quality factors, and provide central access to the current body of knowledge.
\end{abstract}


\section{Introduction}
\label{sec:intro}

\textbf{Context.} A requirements quality factor~\cite{femmer2017thesis} is a normative metric which maps a textual requirement of a specific granularity to a scale which informs about the quality of this input. Because quality factors can be calculated entirely on textual input and do not necessarily need to consider the perspective of any stakeholder who is intended to use the requirement, factors are an efficient tool for early estimates of requirements quality, often even eligible for full automation. This satisfies the need for detecting potential defects in textual requirements at an early stage, as the cost for addressing these defects increases the longer they stay undetected, putting the project success at risk when treated poorly~\cite{fernandez2017naming}. The applicability of quality factors is corroborated by the plethora of existing tools which automate their detection~\cite{montgomery2021empirical, femmer2017rapid}. Among the popular requirements quality factors are \textit{passive voice}~\cite{femmer2017rapid}, where the use of a verb in passive voice is associated with ambiguity of a requirement due to the omission of the subject within a sentence, and \textit{sentence length}~\cite{ferrari2018detecting}, where exceeding a specific threshold of words or characters in a sentence is associated with complexity due to the sentence becoming increasingly hard to comprehend.

\textbf{Problem.} Requirements quality research is lacking a holistic perspective on quality factors and a central repository containing the existing body of knowledge to enable reuse and evolution. These two gaps result in challenges such as concurrent work on same or similar quality factors instead of reusing and advancing those already established. For example, \textit{anaphora} or \textit{anaphoric ambiguity} is described as ``an expression used, in language, to refer to another expression''~\cite{yang2010extending}, ``a linguistic expression that refers to a preceding utterance in text''~\cite{yang2011analysing}, and ``whenever a pronoun (e.g., he, it, that, this, which, etc.) refers to a previous part of the text''~\cite{ferrari2018detecting}. While a certain degree of similarity between all three competing descriptions is apparent, the lack of consensus on the definition is bound to introduce ambiguity to the understanding of the quality factor. Furthermore, another challenge of requirements quality research is the proposal of shallow quality factors neglecting practical relevance due to insufficient or anecdotal evidence~\cite{montgomery2021empirical}. 

\textbf{Approach.} We take the first step at tackling these problems by defining requirements quality factors, their related elements like data sets and automatic detection approaches, and the relationships between the elements. These elements and their relationship constitute our \textit{domain} of interest. Next, we formalize this domain into an \textit{ontology} where each element is represented by an individual \textit{taxonomy} initially derived from literature. Using a set of 105 primary studies from the area of empirical research on requirements quality~\cite{montgomery2021empirical}, we rigorously improve the structure of the ontology by applying established guidelines~\cite{nickerson2013method} and extracted eligible \textit{objects} to populate the ontology. Finally, the refined structure of the ontology as well as all extracted objects are stored centrally in a repository and visualized through a connected web interface.

\textbf{Structure.} After discussing related work in \autoref{sec:related}, we elaborate the long-term objectives of this research direction in \autoref{sec:objective} by explaining the domain of requirements quality factors as well as the corresponding ontology development method in \autoref{sec:design}. \autoref{sec:prototype} describes the first step taken towards this long-term objective in the scope of this work by presenting the process and results of the first prototype at ontology development. Challenges are outlined in \autoref{sec:threats} before calling for action in \autoref{sec:limitations} to involve the requirements quality research community in a joint effort at advancing and maintaining a harmonized vision of requirements quality before concluding in \autoref{sec:conclusion}.

\section{Related Work}
\label{sec:related}
The concept of requirements quality factors has been implicitly used in many publications over the last years: Femmer et al.~\cite{femmer2017rapid}, for instance, introduce nine \textit{requirements smells}, which indicate quality violations in textual requirements. Din and Rine~\cite{din2008requirements} propose a metric for requirements complexity, which is referred to as a \textit{requirements indicator}. Ormandjieva et al.~\cite{ormandjieva2007toward} gather several \textit{quality characteristics} to define the quality of requirements text. We continue using the term \textit{quality factor} which was applied in this context by Femmer et al.~\cite{femmer2015activities}, since the term avoids the negative connotation that for example \textit{requirements smell} evokes, opening the concept of quality factors up to also represent positive impacts on requirements quality, and since the term is well-embedded into a larger context of requirements quality~\cite{femmer2015activities}.

Several sets of requirements quality factors have already been proposed in literature, among which are--as previously mentioned--the requirements smells proposed by Femmer et al.~\cite{femmer2017rapid}, the quality user story framework introduced by Lucassen et al.~\cite{lucassen2016improving}, and the framework for quality measurement developed by G{\'e}nova et al.~\cite{genova2013framework}. Previous attempts at establishing a subject-based classification for requirements quality are to the best of our knowledge limited to an approach by Saavedra et al.~\cite{saavedra2013software}, which is, however, on a coarser granularity and elicits only high-level requirements quality aspects like \textit{correctness}, \textit{completeness}, and others. The work most comparable to our approach has been conducted by Femmer et al.~\cite{femmer2017requirements}, where 129 industrial requirements writing rules were classified regarding their eligibility for automation. Our own work differs from theirs in that (1) we aim at integrating quality factors established in peer-reviewed literature instead of in industrial writing rules~\cite{femmer2017requirements} into a holistic ontology, while (2) considering the eligibility of the individual factors for automation only as one of many sub-goals. Further (3), as our endeavour shall lay the groundwork for a long-term community initiative, one main contribution is to publicly disclose all of our results for an effective maintenance and evolution of the ontology by the community.

\section{Long-Term Objective}
\label{sec:objective}

We begin by framing the long-term objective of our initiative. While this objective is out of scope of this paper, it guides the design and implementation of the prototype.

\subsection{Establishing a Requirements Quality Factor Domain}
\label{sec:objective:domain}

\begin{figure}
\vspace{-.2cm} 
    \centering
    \includegraphics[width=0.85\columnwidth]{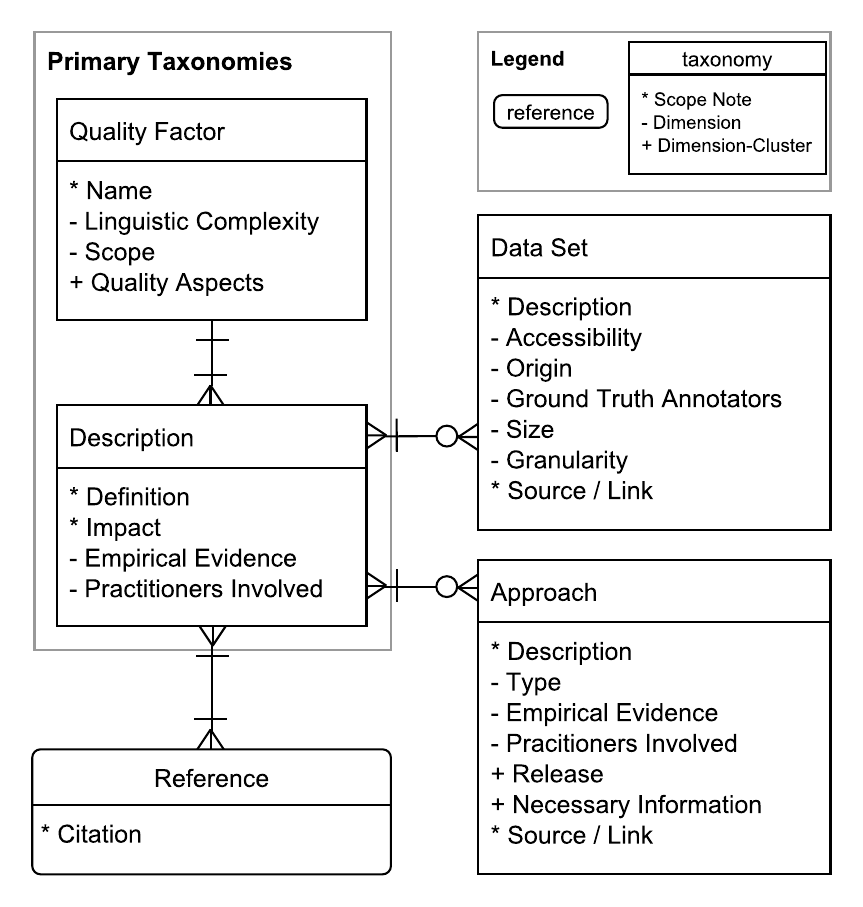}
    \vspace{-.2cm} 
    \caption{Schema of the ontology structure in Crow's foot notation~\cite{everest1976basic}}
    \label{fig:ontology}
    \vspace{-.4cm} 
\end{figure}

Harmonizing the perspectives on requirements quality factors presupposes understanding the domain of related elements in which factors are embedded. We conceptualize four relevant elements during initial investigations of the available literature (see \autoref{fig:ontology}). We consider a \textit{requirements quality factor}--as described in the introduction--as a normative metric which maps a textual requirement at a specific level of granularity to a scale which informs about the quality of this input, where the level of granularity represents different ranges of text (e.g., words, sentences, or documents) and the scale is an often binary categorization of whether the factor has a positive or negative impact on specific aspects of quality (e.g., ambiguity, consistency)~\cite{saavedra2013software}.
The lack of an explicit definition of this concept so far, however, led to quality factors only being referred to implicitly in literature. This resulted in the abstract concept of quality factors being instantiated predominantly as \textit{descriptions} of varying levels of formality in literature. Consequently, the abstract element of a quality factor is related to one or more description elements which define the factor.

Evaluating textual requirements artifacts against these descriptions of quality factors is a way of estimating the quality of the requirements. In several cases, this evaluation can be automated: in our domain, we denote an approach for automatically detecting violations against a quality factor as an \textit{approach}, which is associated to at least one description since approaches often automate the detection of several quality factors. These approaches are evaluated on \textit{data sets}, which may have information about certain quality factors embedded into them, for example through the annotation of violations against a set of quality factors.

We deem these four elements relevant for achieving our objective: the quality factors serve as a conceptual anchor for our objects of interest, descriptions make the factors tangible and comprehensible, and data sets as well as approaches facilitate application and reuse of the factors. We do not claim exhaustive completeness of the domain but rather use it as a starting point for the first iteration. Hence, we focus on extracting these four types of elements from existing literature.

\subsection{Guiding the Ontology Development}
\label{sec:objective:ontology}
In order to formalize these domain elements, we select the simplest subject-based classification system capable of representing the domain elements and their relationships~\cite{garshol2004metadata}. Hence, we formalize our domain of interest as an ontology, where each element is represented by a taxonomy. All \textit{objects} contained in each taxonomy are classified in a fixed number of \textit{dimensions} specific to that taxonomy. As visualized in \autoref{fig:ontology}, an object contained for example in the \textit{quality factor} taxonomy is classified among others by the dimension \textit{scope}. Each object takes exactly one value per dimension, where the set of all possible values of a dimension is called the \textit{characteristics}~\cite{garshol2004metadata}. The dimension \textit{scope} contains the characteristics \textit{word}, \textit{sentence}, and others. We denote the collection of all dimensions and characteristics of a taxonomy as its \textit{structure}. The structure of the ontology is the collection of all taxonomy structures. 

Strictly categorical dimensions are not able to represent certain attributes of an object in the context of our ontology. First, references between two objects of different taxonomies require indexing each object, where indices are not meaningful characteristics. Second, textual attributes like a natural language description of a quality factor can also not be represented by a finite set of characteristics. We therefore extend the attributes of our subject based-classification by indices as well as \textit{scope notes} as commonly used in thesauri~\cite{garshol2004metadata}, which enables a proper description of objects. Each quality factor object for example contains a scope note \textit{name} to associate the object with a unique label.

For the sake of brevity in notation we also introduce \textit{dimension-clusters}, which consist of a list of dimensions and a list of characteristics, where the latter applies to each dimension. A dimension-cluster abbreviates similar dimensions, e.g., the dimension cluster \textit{quality aspects} of the quality factor taxonomy contains dimensions like ambiguity, complexity, and verifiability, which all can take the characteristics \textit{impacted positively}, \textit{impacted negatively}, or \textit{not impacted} individually.

We translate the rigorous taxonomy development guideline proposed by Nickerson et al.~\cite{nickerson2013method} and extended by Kundisch et al.~\cite{kundisch2021update} to the larger scale of our ontology. None of the aspects in which our ontology design extends the design of a taxonomy contradicts the abstract guidelines, since (1) the ontology simply consists of four individual taxonomies and (2) the extraction guideline is applicable to the additionally included index and scope note attribute as well. 

\section{Provisional Ontology Design}
\label{sec:design}

We take a step towards our long-term objective by defining requirements and exit criteria of the ontology creation process.

\subsection{Meta-Characteristics}
\label{sec:design:meta}
As defined in the guideline~\cite{nickerson2013method}, the design of the classification system is rooted in the identification of the users which are intended to use the ontology, and their goals, which these users are supposed to achieve with the ontology. We consider two types of users to interact with the ontology: \textit{researchers} dedicated to advancing the field of requirements quality research and \textit{practitioners} aiming to apply results emerging from this research in order to evaluate the quality of their requirements artifacts. The goals of these two abstract users constitute the meta-characteristics~\cite{nickerson2013method} and represent the high-level requirements for the ontology. The following goals (G) are formulated in the user story template:
\begin{itemize}[nosep,leftmargin=1em,labelwidth=*,align=left]
    \item \textbf{G1:} As a researcher or practitioner, I want to find explanations to available requirements quality factors so that I understand how they inform about requirements quality.
    \item \textbf{G2:} As a researcher or practitioner, I want to find available resources connected to a quality factor so that I can reuse these resources for my own work.
    \item \textbf{G3:} As a researcher, I want to identify gaps in literature so that I can tailor my own research to provide valid contributions.
    \item \textbf{G4:} As a researcher or practitioner, I want to find who is working on specific quality factors so that I can establish a collaboration.
\end{itemize}
These goals strictly apply to the \textit{ontology}. The obvious, overarching goal to evaluate the quality of requirements artifacts applies to the \textit{quality factors} and is independent of our goals. 

\subsection{Exit criteria}
\label{sec:design:exit}
The exit criteria, which indicate the completeness of the iterative ontology development method, are also derived from Nickerson et al.~\cite{nickerson2013method}. In this, we aim to achieve the following objective ending conditions (condensed from~\cite{nickerson2013method}): (1) All objects or a representative sample of objects have been examined, (2) no object, dimension or characteristic was merged or split in the last iteration, (3) at least one object is classified under every characteristics of every dimension, (4) no new dimensions or characteristics were added in the last iteration, and (5) every dimension is unique in every taxonomy of the ontology and every characteristic is unique in its dimension.

By documenting all changes to the ontology in each iteration, we can objectively decide when these ending conditions are met. We have explicitly excluded the objective ending criterion ``Each cell (combination of characteristics) is unique and is not repeated''~\cite{nickerson2013method}, since the extension of the ontology as described in \autoref{sec:objective:ontology} entailed the inclusion of attributes that are not dimensions, i.e., indices and scope notes. Two objects can hence have the same combination of characteristics among all dimensions, but be distinct due to different scope note values. In addition, we also aim to achieve the following subjective ending conditions~\cite{nickerson2013method}:
\begin{itemize}[nosep,leftmargin=1em,labelwidth=*,align=left]
    \item \textbf{Concise}: the number of dimensions needs to be meaningful yet manageable
    \item \textbf{Robust}: the dimensions and characteristics need to provide for differentiation among objects of interest
    \item \textbf{Comprehensive}: all objects within the domain of interest can be classified
    \item \textbf{Extendable}: new dimensions and characteristics can be easily added
    \item \textbf{Explanatory}: the dimensions and characteristics explain the objects
\end{itemize}

\section{Prototype of the Ontology}
\label{sec:prototype}

We developed a prototype of the ontology to (1) illustrate the usability of the structure, repository, and tool, and to (2) contribute the first step towards the long-term objectives.

\subsection{Iterative process}
\label{sec:prototype:process}
We illustrate the approach outlined in \autoref{sec:objective:ontology} by adopting the iterative ontology development process as extended from Nickerson et al.~\cite{nickerson2013method}. Accordingly, either an empirical-to-conceptual or conceptual-to-empirical approach has to be chosen. We chose the former approach for the initial iteration, as a significant understanding of the domain has already been established along previous engagement with requirements quality research. We distilled an initial structure of the ontology based on relevant literature~\cite{unterkalmsteiner2017requirements,femmer2017requirements,saavedra2013software}.

An empirical-to-conceptual approach was chosen for the subsequent four iterations, as we aim to extract eligible objects for the four taxonomies from established literature. For this prototype, we selected the set of primary studies gathered in a recent systematic mapping study on empirical requirements quality research by Montgomery et al.~\cite{montgomery2021empirical} as our data to extract from. This publication is the only secondary study to our knowledge which explicitly investigates requirements quality and, thus, serves as a reliable collection of peer-reviewed primary studies. The three first authors distributed the set of 105 eligible studies among each other and split the resulting subset into four iterations. During each iteration, they extracted all relevant objects based on an extraction guideline, which was initiated during the first iteration and maintained according to ontology development protocol~\cite{nickerson2013method,kundisch2021update}. Publications had to at least contain one eligible quality factor based on the definition established in \autoref{sec:intro} and an according description. Data set and approach objects were extracted when eligible according to the extraction guideline. At the end of each iteration, the three extracting authors convened together with the fourth author and discussed necessary changes to the ontology structure in case objects were encountered which could currently not be framed by the taxonomies.

The set of references in~\cite{montgomery2021empirical} is heavily biased towards empirical work. To confirm that the ontology is also robust when considering non-empirical work, we conducted a final iteration considering publications that were excluded in the reference selection phase of~\cite{montgomery2021empirical}. The inclusion of non-empirical work, e.g.,~\cite{fabbrini1998achieving}, did not challenge the structure of any taxonomy, strengthening our confidence in the robustness of the ontology.

After this final iteration, all relevant exit criteria were fulfilled, which indicated the completion of the ontology creation process in the scope of this work. The objective ending conditions were fulfilled as the documentation of the final iteration of the protocol showed no violation against any of the five conditions. The subjective ending conditions were assessed and agreed upon by the first four authors to a reasonable extent of this prototype; for example, the ontology was deemed \textit{concise} since the number of dimensions of each taxonomy is compliant with the seven plus two rule \cite{nickerson2013method, miller1956magical}, and \textit{robust} since the inclusion of non-empirical work did not challenge any taxonomy structure. The final assessment of the subjective ending conditions applies to the future version of the ontology and will be discussed in the outlook in \autoref{sec:limitations:call}.

\subsection{Current State}
\label{sec:prototype:state}
The schema of the ontology structure at the current stage of development is shown in \autoref{fig:ontology} with the structure and relationship between all four included taxonomies. A thorough explanation of all attributes (dimensions, dimension-clusters, scope notes, and references), the eligible characteristics, and their corresponding extraction rule can be found in our replication package\footnote{\label{fn:replication}Replication package at \url{https://doi.org/10.5281/zenodo.6583690}} and on our web interface\footnote{\label{fn:tool}The application can be accessed at \url{http://www.reqfactoront.com}}. We limit the following explanations to the most important of these attributes and illustrate them with a running example.

Requirements quality factors are characterized by a \textit{name} and \textit{scope}, which is the dimension representing the granularity of input necessary in order to decide the quality factor. As an example, one publication by Femmer et al.~\cite{femmer2017requirements} contains multiple quality factors, one of which is named \textit{containing subflows}. The scope of this factor is \textit{use case}, as a full use case is necessary to decide whether at least one subflow is contained or not. Quality factors are further characterized by the dimension-cluster quality \textit{aspect}: it is necessary to denote the impact which the calculated value of a quality factor has on the activities in which the requirement is used, which is framed by the notion of activity-based requirements quality~\cite{femmer2015activities}. The factor \textit{containing subflows} is reported to have a \textit{negative} effect on the aspect \textit{understandability}, because subflows ``force the reader to jump between different positions in the text in order to read through the use case, which can be argued to lead to less readable use cases that are harder to understand''~\cite{femmer2017requirements}, but also a \textit{positive} effect on \textit{maintainability}, because ``if parts of the flow change, they only need to be changed in one location (the subflow), and not in each use case''~\cite{femmer2017requirements}. The notion of aspects is further explained in \autoref{sec:threats}. The set of quality aspects is a harmonized superset of aspects used in established literature~\cite{schneider2002process, montgomery2021empirical, ISO29148, saavedra2013software}. We make no claim about the completeness and granularity of this set, as we consider quality aspects as a connected, but distinct element in a larger domain. Using a harmonized superset, we provide an interface for future research in this subsequent domain of requirements quality, for example, on their interrelationships~\cite{unterkalmsteiner2017requirements}.

Descriptions are instantiated by a scope note for both the \textit{definition} of the quality factor and also its \textit{impact}. Since a rigorous framework of requirements quality factors has been absent in requirements quality research, textual descriptions of what a quality factor means and how it impacts subsequent development activities are most common. The factor \textit{containing subflows} is defined as ``Subflows are mechanisms for reuse that enable the author of a use case to extract a certain set of steps into a reusable subflow to prevent copy-and-paste reuse [...] in the use cases''~\cite{femmer2017rapid}, while the impact is described as mentioned in the previous paragraph about aspects. Description objects are further annotated on whether the according publication provides \textit{empirical evidence} for its relevance and whether \textit{practitioners were involved} in its inception or development, as these dimensions help identifying quality factors that are empirically informed. Since all quality factors in~\cite{femmer2017requirements} were derived from an industrial requirements writing guideline, they explicitly have \textit{practitioners involved} and their use in practice serves as \textit{empirical evidence}.

Data sets are characterized by their \textit{origin}, which reflects whether the data is from industry, academia, or mocked, and who embedded the information (called \textit{ground-truth annotators}) of quality factor violations in the data, i.e., whether the authors themselves, practitioners, or students annotated the violations. Femmer et al.~\cite{femmer2017requirements} report on one data set from a large software project at a German reinsurance company, whose \textit{origin} is \textit{practitioner data}. Since the data bears no annotations, the data set has no ground-truth annotators. The \textit{size} and \textit{granularity} of the data set represents the number and type of contained elements. The aforementioned data set contains 51 objects of the granularity \textit{document}. Finally, the \textit{accessibility} of a data set reflects to what degree the data set can be used. If available, the corresponding link or reference to the source is given. The described data set is \textit{private} and has no link or source given.

A similar approach for characterizing the \textit{accessibility} is done for an approach object, in addition to the type of \textit{release} (source code, tool, API, or other). Approaches are further characterized by their \textit{type} (rule-based, machine learning or deep learning) and the \textit{necessary information} utilized to conduct the automatic evaluation (e.g., POS tags, dependency tags, or other). Finally, approaches are--similar to descriptions--classified regarding the \textit{empirical evidence} they provide and whether \textit{practitioners were involved} in the evaluation. The approach \textit{Smella}, described in another publication by Femmer et al.~\cite{femmer2017rapid}, is a \textit{proprietary tool} detecting smells with a \textit{rule-based} algorithm using \textit{POS tags} and \textit{lemmatization}.

In the following paragraphs we describe how the current state of the ontology and its contained objects address the meta-characteristics. All conclusions are drawn based on the limited subset that was selected for this prototype~\cite{montgomery2021empirical}, hence the inferences are not necessarily universal. In addition, the conclusions are currently limited to a quantitative evaluation of the ontology's structure and content. A qualitative evaluation involving the intended users of the ontology will be necessary to determine its usability.

\textbf{Addressing G1.} The association of a quality factor object with at least one description object provides an overview over all proposed explanations of a quality factor. Out of the 105 primary studies from the initial set~\cite{montgomery2021empirical}, 59 contained at least one eligible quality factor. In total, 206 unique quality factors were extracted and associated to 258 descriptions. Consequently, 172 quality factors are associated to exactly one description. On the other hand, nine quality factors were described in three or more occasions: anaphora, coordination ambiguity, vagueness, passive voice, referential integrity, subjectivity, nocuous ambiguity, multiple interpretations, and consistency. 

\textbf{Addressing G2.} The association of a quality factor object with data set and approach objects allows to find available resources for reuse and evolution. The 105 primary studies describe 56 unique data sets and 36 approaches. However, only 9 of 56 data sets are publicly available (i.e., have the characteristic \textit{available in paper} or \textit{open access link} in the dimension \textit{accessibility}), while most data sets are either private or not disclosed. Only 5 of the 36 approaches are publicly available (i.e., have the characteristic \textit{open access} or \textit{open source} in the dimension \textit{accessibility}), while most approaches are not disclosed at all. These numbers highlight a dire condition of open source in the requirements quality research landscape, which inhibits the use and reuse of existing resources. Filtering by the dimension \textit{accessibility} supports identifying available resources and exposing undisclosed contributions.

\textbf{Addressing G3.} As well as achieving G2 in the aforementioned way to identify which quality factors are not yet annotated in a data set or automatically detected with an approach implementation, the dimensions \textit{empirical evidence} and \textit{practitioners involved} can be used as a filter to identify objects that lack empirical validation. Out of the 258 descriptions, only 82 are devised based either on empirical evidence, i.e., by assessing how well the metrics correspond to the subjective perception of requirements quality in a survey~\cite{usdadiya2018assessing}, or by involving practitioners in the design process of the quality factor~\cite{antinyan2016complexity}. In addition, only 92 of the extracted 258 description objects contain an explicit \textit{impact} of the quality factor. The significance of this lack of an impact description is further emphasized in \autoref{sec:threats}.

\textbf{Addressing G4.} The association of description, data set, and approach object to references allows to trace every contribution to the corresponding authors, which can be used to connect to researchers who have contributed in a specific area of requirements quality research.


\subsection{Repository and Tool}
\label{sec:prototype:repo}
The initial results are recorded in a first version of a maintainable tool: both the structure and the objects of the ontology are stored in a publicly accessible data repository hosted on GitHub\textsuperscript{\ref{fn:replication}}. The structure is represented by structure files defining the attributes of each taxonomy. The objects are stored in form of \textit{extractions}, where each extraction is associated to one reference and contains an arbitrary number of extracted objects according to the existing taxonomies. The current status of the repository is retrieved by an interactive web application which processes the data and visualizes it in a human-readable and -comprehensible way\textsuperscript{\ref{fn:tool}}, fulfilling the elicited goals through filters and links. The repository can be easily maintained using the version control offered by GitHub: contributions to both the structure and the content of the ontology can be made by adding new extraction elements for either existing or new references. This way, new publications can be included or already included publications can be revised, supporting an inclusive and collaborative approach at harmonizing the perspective on requirements quality factors.

\section{Threats and Challenges}
\label{sec:threats}

\paragraph{Transparent Ontology Design Process}
A major challenge in developing any subject-based classification is the lack of transparency of the process~\cite{kundisch2021update}, where the process obscures the rationale behind design decisions. Since our ontology is both meant to facilitate collaboration and a community-driven maintenance and evolution, we mitigate this threat by disclosing all process documentations\textsuperscript{\ref{fn:replication}}.

\paragraph{Shared Understanding of Extraction Guidelines}
As with any extraction task, the subjective nature of interpreting literature according to an extraction guideline is inherently prone to misunderstandings. Even though the initial set of extracted objects are neither the main contribution of this work nor assumed to be permanent, we assured a common understanding by assigning primary studies which were already processed by one author to another author in order to calculate an overlap and quantify the agreement. This way, each of the first three authors additionally extracted relevant objects from two already processed primary studies, such that every extractor had an overlap with every other extractor. All relevant attributes were evaluated: dimensions were assessed by equivalence, scope notes were assessed by similarity using sequence matching scaled to range [0, 1]. The six primary studies resulted in 799 extracted individual values, on which an agreement of 85.03\% between all authors was achieved. This agreement assures a sufficiently common understanding of the extraction guidelines.

\paragraph{Requirements Quality Research Framework}
As mentioned in \autoref{sec:intro}, requirements quality factors are purely normative and evaluate textual input based on metrics which are often arbitrary. The relationship between these metrics and the actual impact on the quality is more complex: as identified in previous research on specific quality factors~\cite{femmer2014impact}, a violation against the rule entailed by a quality factor may or may not lead to an actual impact on the requirements quality depending on numerous context factors. For example, the use of passive voice might not lead to an ambiguous interpretation in a small-scale development unit if the stakeholders which are intended to use the written requirement can reconstruct the omitted subject of the sentence anyway. 

This relationship has been framed by Femmer et al. in the form of activity-based requirements engineering quality models~\cite{femmer2015activities}, where a violation against a quality factor only potentially leads to an impact on an activity in which the requirement is meant to be used. The relevance of a quality factor is dependent on the likelihood of an impact on subsequent activities under the given context factors.

This imposes a necessary interface on the requirements quality factor ontology: ultimately, every quality factor should be associated to a specific impact on specific activities given specific context factors in order to determine the relevance of the factor. The state of research in this respect is currently relatively poor, as shown in the preliminary results of \autoref{sec:prototype:state}, and most publications proposing quality factors are satisfied with determining the impact of a factor based on educated guesses or anecdotal evidence. Therefore, our ontology currently only records explicitly stated impacts in the dimension-cluster \textit{quality aspect} of the taxonomy \textit{quality factor}. However, improving the information about the potential impact of quality factors is an anticipated extension point of our ontology once research in this domain has advanced.

\section{Limitations and Call for Action}
\label{sec:limitations}

We discuss the limitations of the prototype to highlight the distance between this first step and the long-term objective. Further, we propose a community effort to bridge this distance.

\subsection{Limitations of the current Approach}
\label{sec:limitations:limitations}

\textbf{Incompleteness of Publications} The lack of a shared terminology impedes identifying what the complete set of publications would be, as quality factors have been addressed with different names and in different approaches. Hence, the list of publications to extract eligible objects from is far from complete. The systematic mapping study on empirical requirements quality research by Montgomery et al.~\cite{montgomery2021empirical} is to our knowledge the only secondary study which makes an attempt at comprehending the research domain of requirements quality. Currently not considered publications could potentially add relevant objects to the ontology or challenge its structure. Since the domain of requirements quality research is only loosely coherent by an explicit identity, the effort to comprehend and order relevant research is, as we argue, an extensive undertaking.

\textbf{Overload of factors} The result addressing goal G1 presented in \autoref{sec:prototype:state} raises the question about the relevance of this large number of unique quality factors. The included publications show a large variation in the degree of evidence for their relevance, as also noticed by Montgomery et al.~\cite{montgomery2021empirical}, which ranged from purely anecdotal justifications over references to established literature~\cite{juergens2010can} to sound empirical evaluations~\cite{femmer2014impact}. We decided not to exclude publications with lacking evidence of relevance at the cost of a manageable number of resulting factors, mainly because no mature research approach to reliably determine a quality factor's relevance exists yet.


\subsection{Call for Action}
\label{sec:limitations:call}

One hope we associate with this RE@Next! contribution is to appeal for participation in a coordinated community effort aimed at tackling this task. The extension of this task to a community effort makes the extensive undertaking of identifying all relevant literature surmountable. In addition, it ensures to include diverse perspectives on the matter, contributing to establish a harmonized vision. This will additionally lead to healthy scrutiny and subsequent evolution of the ontology structure, for example by including the dimension \textit{language} for quality factors, as publications discussing quality factors in languages other than English begin to emerge~\cite{hasso2019detection}. Finally, involving as many parts of the implicit requirements quality research community as possible is bound to establish an explicit, shared identity of the research domain in the process.

The community effort will be initialized by interested members of the requirements quality research community committing to it. We anticipate this effort to span over several years, though a consistent commitment is not mandatory. Coordinated by the first authors of this paper, systematic strategies for identifying previously not considered publications will be developed, distributed, and executed. Once confidence in the completeness of the publications will have been reached, the iterative ontology creation process described in \autoref{sec:prototype:process} will be scaled up and continued by the members involved in the community effort. The ultimate deliverable of this community effort will be a sufficiently complete and robust ontology structure and content--assessed jointly using the objective and subjective exit criteria--which reflects the harmonized perspectives of the requirements quality research community.

This also lays the groundwork for addressing the relevance-problem of requirements quality publications: after the space of quality factors has expanded during the community effort, this same community shall be involved in developing a reliable research approach for determining the relevance of a quality factor. This method will be used to condense the space of quality factors again to a manageable number of relevant objects, addressing the second limitation mentioned in the \autoref{sec:limitations:limitations}. Finally, a complete yet concise set of applicable and relevant quality factors contained in the final version of the ontology fulfilling goals G1-G4 can be delivered.

\section{Conclusion and Outlook}
\label{sec:conclusion}

This paper presents the long-term objective of a harmonized vision on requirements quality factors in the form of an ontology, relating four taxonomies to represent the four elements quality factor, description, data set, and approach of the domain containing quality factors for textual requirements. The extraction of eligible objects from 105 primary studies as well as a central repository and accessible web interface are the first step towards this long-term objective.

Establishing a harmonized perspective on the structure of quality factors and related elements as well as a central repository containing a sufficiently complete set of relevant objects is an extensive task necessitating a community effort, making this task surmountable and also including diverse perspectives on the domain. The final version of the ontology will then serve as a conceptual framework for future research, a reliable resource for practitioners to base requirements quality assurance on, and a tool for requirements quality education. 

\section*{Acknowledgments}
This work was supported by the KKS foundation through the S.E.R.T. Research Profile project at Blekinge Institute of Technology.

\bibliographystyle{IEEEtran}
\bibliography{material/references}

\begin{thebibliography}{10}
\providecommand{\url}[1]{#1}
\csname url@samestyle\endcsname
\providecommand{\newblock}{\relax}
\providecommand{\bibinfo}[2]{#2}
\providecommand{\BIBentrySTDinterwordspacing}{\spaceskip=0pt\relax}
\providecommand{\BIBentryALTinterwordstretchfactor}{4}
\providecommand{\BIBentryALTinterwordspacing}{\spaceskip=\fontdimen2\font plus
\BIBentryALTinterwordstretchfactor\fontdimen3\font minus
  \fontdimen4\font\relax}
\providecommand{\BIBforeignlanguage}[2]{{%
\expandafter\ifx\csname l@#1\endcsname\relax
\typeout{** WARNING: IEEEtran.bst: No hyphenation pattern has been}%
\typeout{** loaded for the language `#1'. Using the pattern for}%
\typeout{** the default language instead.}%
\else
\language=\csname l@#1\endcsname
\fi
#2}}
\providecommand{\BIBdecl}{\relax}
\BIBdecl

\bibitem{femmer2017thesis}
H.~Femmer, ``Requirements engineering artifact quality: definition and
  control,'' Ph.D. dissertation, Technische Universit{\"a}t M{\"u}nchen, 2017.

\bibitem{fernandez2017naming}
D.~M{\'e}ndez~Fern{\'a}ndez, S.~Wagner, M.~Kalinowski, M.~Felderer, P.~Mafra,
  A.~Vetr{\`o}, T.~Conte, M.-T. Christiansson, D.~Greer, C.~Lassenius
  \emph{et~al.}, ``Naming the pain in requirements engineering: Contemporary
  problems, causes, and effects in practice,'' \emph{EMSE}, vol.~22, no.~5,
  2017.

\bibitem{montgomery2021empirical}
L.~Montgomery, D.~Fucci, A.~Bouraffa, L.~Scholz, and W.~Maalej, ``Empirical
  research on requirements quality: A systematic mapping study,'' \emph{REJ},
  2021.

\bibitem{femmer2017rapid}
H.~Femmer, D.~M. Fern{\'a}ndez, S.~Wagner, and S.~Eder, ``Rapid quality
  assurance with requirements smells,'' \emph{JSS}, vol. 123, 2017.

\bibitem{ferrari2018detecting}
A.~Ferrari, G.~Gori, B.~Rosadini, I.~Trotta, S.~Bacherini, A.~Fantechi, and
  S.~Gnesi, ``Detecting requirements defects with nlp patterns: an industrial
  experience in the railway domain,'' \emph{EMSE}, vol.~23, no.~6, 2018.

\bibitem{yang2010extending}
H.~Yang, A.~De~Roeck, V.~Gervasi, A.~Willis, and B.~Nuseibeh, ``Extending
  nocuous ambiguity analysis for anaphora in natural language requirements,''
  in \emph{RE}, 2010.

\bibitem{yang2011analysing}
------, ``Analysing anaphoric ambiguity in natural language requirements,''
  \emph{REJ}, vol.~16, no.~3, 2011.

\bibitem{nickerson2013method}
R.~C. Nickerson, U.~Varshney, and J.~Muntermann, ``A method for taxonomy
  development and its application in information systems,'' \emph{European
  Journal of Information Systems}, vol.~22, no.~3, 2013.

\bibitem{din2008requirements}
C.~Y. Din and D.~Rine, \emph{Requirements content goodness and complexity
  measurement based on NP chunks}.\hskip 1em plus 0.5em minus 0.4em\relax VDM
  Publishing, 2008.

\bibitem{ormandjieva2007toward}
O.~Ormandjieva, I.~Hussain, and L.~Kosseim, ``Toward a text classification
  system for the quality assessment of software requirements written in natural
  language,'' in \emph{SOQUA}, 2007.

\bibitem{femmer2015activities}
H.~Femmer, J.~Mund, and D.~M. Fern{\'a}ndez, ``It's the activities, stupid! a
  new perspective on re quality,'' in \emph{RET}, 2015.

\bibitem{lucassen2016improving}
G.~Lucassen, F.~Dalpiaz, J.~M.~E. van~der Werf, and S.~Brinkkemper, ``Improving
  agile requirements: the quality user story framework and tool,''
  \emph{Requirements engineering}, vol.~21, no.~3, pp. 383--403, 2016.

\bibitem{genova2013framework}
G.~G{\'e}nova, J.~M. Fuentes, J.~Llorens, O.~Hurtado, and V.~Moreno, ``A
  framework to measure and improve the quality of textual requirements,''
  \emph{Requirements engineering}, vol.~18, no.~1, pp. 25--41, 2013.

\bibitem{saavedra2013software}
R.~Saavedra, L.~C. Ballejos, and M.~A. Ale, ``Software requirements quality
  evaluation: State of the art and research challenges,'' in \emph{ASSE-JAIIO},
  2013.

\bibitem{femmer2017requirements}
H.~Femmer, M.~Unterkalmsteiner, and T.~Gorschek, ``Which requirements artifact
  quality defects are automatically detectable? a case study,'' in \emph{REW},
  2017.

\bibitem{everest1976basic}
G.~C. Everest, ``Basic data structure models explained with a common example,''
  in \emph{Proc. Fifth Texas Conference on Computing Systems}, 1976.

\bibitem{garshol2004metadata}
L.~M. Garshol, ``Metadata? thesauri? taxonomies? topic maps! making sense of it
  all,'' \emph{Journal of information science}, vol.~30, no.~4, 2004.

\bibitem{kundisch2021update}
D.~Kundisch, J.~Muntermann, A.~M. Oberl{\"a}nder, D.~Rau, M.~R{\"o}glinger,
  T.~Schoormann, and D.~Szopinski, ``An update for taxonomy designers,''
  \emph{Business \& Information Systems Engineering}, pp. 1--19, 2021.

\bibitem{unterkalmsteiner2017requirements}
M.~Unterkalmsteiner and T.~Gorschek, ``Requirements quality assurance in
  industry: why, what and how?'' in \emph{REFSQ}.\hskip 1em plus 0.5em minus
  0.4em\relax Springer, 2017.

\bibitem{fabbrini1998achieving}
F.~Fabbrini, M.~Fusani, V.~Gervasi, S.~Gnesi, and S.~Ruggieri, ``Achieving
  quality in natural language requirements,'' in \emph{QW}, 1998.

\bibitem{miller1956magical}
G.~A. Miller, ``The magical number seven, plus or minus two: Some limits on our
  capacity for processing information.'' \emph{Psychological review}, vol.~63,
  no.~2, p.~81, 1956.

\bibitem{schneider2002process}
R.~E. Schneider, \emph{A process for building a more effective set of
  requirement goodness properties}.\hskip 1em plus 0.5em minus 0.4em\relax
  George Mason University, 2002.

\bibitem{ISO29148}
``{Systems and software engineering — Life cycle processes — Requirements
  engineering},'' International Organization for Standardization, Geneva, CH,
  Standard, Nov. 2018.

\bibitem{usdadiya2018assessing}
C.~P. Usdadiya, ``Assessing quality of use case specifications,'' Ph.D.
  dissertation, Dhirubhai Ambani Institute of Information and Communication
  Technology, 2018.

\bibitem{antinyan2016complexity}
V.~Antinyan, M.~Staron, A.~Sandberg, and J.~Hansson, ``A complexity measure for
  textual requirements,'' in \emph{IWSM-MENSURA}, 2016.

\bibitem{femmer2014impact}
H.~Femmer, J.~Ku{\v{c}}era, and A.~Vetr{\`o}, ``On the impact of passive voice
  requirements on domain modelling,'' in \emph{ESEM}, 2014.

\bibitem{juergens2010can}
E.~Juergens, F.~Deissenboeck, M.~Feilkas, B.~Hummel, B.~Schaetz, S.~Wagner,
  C.~Domann, and J.~Streit, ``Can clone detection support quality assessments
  of requirements specifications?'' in \emph{ICSE}, 2010, pp. 79--88.

\bibitem{hasso2019detection}
H.~Hasso, M.~Dembach, H.~Geppert, and D.~Toews, ``Detection of defective
  requirements using rule-based scripts.'' in \emph{REFSQ Workshops}, 2019.

\end{thebibliography}

\end{document}